\documentclass[%
 reprint,
 amsmath,amssymb,
 aps,
]{revtex4-2}

\usepackage{graphicx}
\usepackage{dcolumn}
\usepackage{xcolor}
\usepackage{bm}

\newcommand{\RB}[1]{\textcolor{black}{#1}}
\newcommand{\RBB}[1]{\textcolor{black}{#1}}

\newcommand{\LC}[1]{\textcolor{black}{#1}}
\renewcommand{\phi}{\varphi}
\renewcommand{\epsilon}{\varepsilon}

\begin{document}


\title{Decoupling Structure and Elasticity in Colloidal Gels Under Isotropic Compression}

\newcommand{\equalcontrib}{These authors contributed equally to this work.}

\author{M. Milani$^1$}
\email{matteo.milani@espci.fr}
\altaffiliation[Current affiliation: PMMH ESPCI, Paris, France]{}
\author{E. Cavalletti$^1$}%
\author{V. Ruzzi$^1$}%
\author{A. Martinelli$^1$}%
\author{P. Dieudonné-George$^1$}%
\author{C. Ligoure$^1$}%
\author{T. Phou$^1$}%
\author{L. Cipelletti$^{1,2}$}%
\altaffiliation{These authors contributed equally.}

\author{L. Ramos$^{1,\dag,}$}%
\email{laurence.ramos@umontpellier.fr}

\affiliation{%
$^1$Laboratoire Charles Coulomb (L2C), Universit\'e Montpellier, CNRS, Montpellier, France \\
$^2$Institut Universitaire de France, Paris, France
}%

\date{\today}

\begin{abstract}
We exploit the controlled drying of millimeter-sized gel beads to investigate isotropic compression of colloidal fractal gels. Using a custom dynamic light scattering setup, we demonstrate that stresses imposed by drying on the bead surface propagate homogeneously throughout the gel volume, inducing plastic rearrangements. We find that the Young modulus and yield stress of the gels increase monotonically with the instantaneous colloid volume fraction, $\phi$, exhibiting a mechanical response that depends solely on $\phi$, regardless of the drying history. In striking contrast, small-angle X-ray scattering reveals that the gel microstructure retains a strong memory of its initial state, depending on both $\varphi$ and the entire compression pathway. Our findings challenge the prevailing paradigm of a one-to-one relationship between microstructure and elasticity in colloidal fractal gels, opening new avenues for independent control over the structural and mechanical properties of soft materials.
\end{abstract}

\maketitle


Colloidal gels, ubiquitous in applications from food to ceramics, are viscoelastic solids formed by a percolating network of aggregated particles immersed in a liquid~\cite{guine2018drying,van2009colloidal,xing2016colloid,augier2002risk,omatete1991drying,chiu1993drying,colina2000drying}. Their elasticity, tunable over orders of magnitude by parameters like particle interactions and volume fraction, $\phi$,~\cite{shih1990scaling,wyss2004small,ruiz-franco_tuning_2020,nikoumanesh_elucidating_2024,buscall_scaling_1988,rueb1997viscoelastic} has made them model systems for amorphous solids. Despite significant theoretical and numerical advancements~\cite{shih1990scaling,bantawa2023hidden}, the subtle links between microstructure and viscoelasticity remain elusive. Crucially, a definitive experimental demonstration explicitly connecting macroscopic mechanical properties to microscopic structural features is still lacking, with many studies relying on theoretical predictions from over three decades~\cite{shih1990scaling}.

A way to fine-tune $\phi$ and thus gel elasticity is compression~\cite{wyss_mechanics_2005}. Numerical studies have explored uniaxial compression of strong colloidal gels, predicting consolidation and universal scaling of yield stress with $\phi$~\cite{roy2016universality,seto2013compressive,islam2021consolidation}. However, experimental data on compressed gels are scarce~\cite{wyss_mechanics_2005}. Most experiments relating structure and rheology often rely on gravity-induced gel collapse~\cite{manley2005gravitational, brambilla2011highly, tateno_void_2025}, which limits precise $\phi$ tuning and simultaneous measurement of mechanical and structural characteristics. Such methods also suffer from \LC{wall interactions}~\cite{starrs_collapse_2002} \LC{and potentially anisotropic particle distributions}~\cite{islam2021consolidation}.  Alternatively, drying offers multidirectional compaction~\cite{brown2002consolidation,islam2021consolidation}, but frequently causes crack formation~\cite{lakshmikantha2009image,chiu1993drying,kindle1917some,goehring2010evolution}, bending~\cite{ogawa2020drying}, or delamination~\cite{sarkar2011delamination}. However, recent work has shown crack-free drying of colloidal gel slabs when substrate adhesion is suppressed~\cite{thiery2015water,thiery2016drying}.

Here, we present a multiscale investigation of colloidal gels under isotropic compression, leveraging the smooth, crack-free drying of boundary-free millimeter-sized gel beads. This approach allows us to precisely tune $\phi$ over a wide range while simultaneously probing gel elasticity and microstructure. By comparing gels dried to the same $\phi$ but originating from different initial volume fractions ($\phi_0$), we decouple the roles of instantaneous concentration and compression history. Our results reveal a surprising breakdown of the commonly accepted one-to-one relationship between microstructure and elasticity, a cornerstone of fractal colloidal gels, thereby challenging recent numerical predictions of a universal framework independent of sample preparation~\cite{Richard2025}. This discovery unveils a new phase space for colloidal gels, where mechanical and structural properties can be independently controlled.

Millimetric gel beads are produced using the protocol of~\cite{milani2025synthesis}. Silica nanoparticles (NPs, Ludox AS-30 from Sigma Aldrich, diameter 6 nm) are suspended at initial volume fractions $\varphi_0 \in [1.5\%-14\%]$ in an aqueous solvent that comprises 1M urea and the enzyme urease (U 5378) from Canavalia ensiformis (from Sigma-Aldrich), at a fixed concentration of $\approx 30$ U/mL. We deposit a $30$ $\mu$L drop of NP suspension on a hydrophobic surface and immerse it in a bath of silicon oil with a density lower than that of the suspension. In our experimental conditions, the capillary length is of the order of  $13$ mm, much larger than the drop size, ensuring that the drop remains spherical due to surface tension~\cite{milani2025synthesis}. With time, the enzymatic reaction triggers NPs aggregation, leading to gelation~\cite{wyss2004small, aime2018power}. The gel beads thus formed have an initial radius $R_0=2$ mm; they are kept in oil for at least $24$ h to ensure that gelation and aging are complete. To investigate the compression process, we remove the bead from the oil bath, thoroughly rinse it with cyclohexane, and deposit it immediately on a hydrophobic surface, where it undergoes an isotropic, crack-free drying process~\cite{milani2025synthesis}. To vary the compression rate defined as $\dot{\varepsilon_v}=\frac{\dot{R}}{R}$, with $R$ the instantaneous radius of the bead, the bead is enclosed in a glass box with regulated relative humidity (RH)~\cite{milani2023double}, which typically ranges between $20\%$ and $94\%$. The beads undergo a two-step drying process~\cite{milani2025synthesis,simpkins1989drying,brinker2013sol}: initially, the beads shrink smoothly, leading to an increase in the NPs' volume fraction, $\varphi$. \LC{Later, $R$ attains a limiting value and air invades the gel}. Here, we focus on the structural and mechanical modifications of the gel occurring in the first drying regime, when the colloidal gel is compressed isotropically and remains fully wet.

\begin{figure}[htbp] 
\includegraphics[width=0.8\columnwidth]{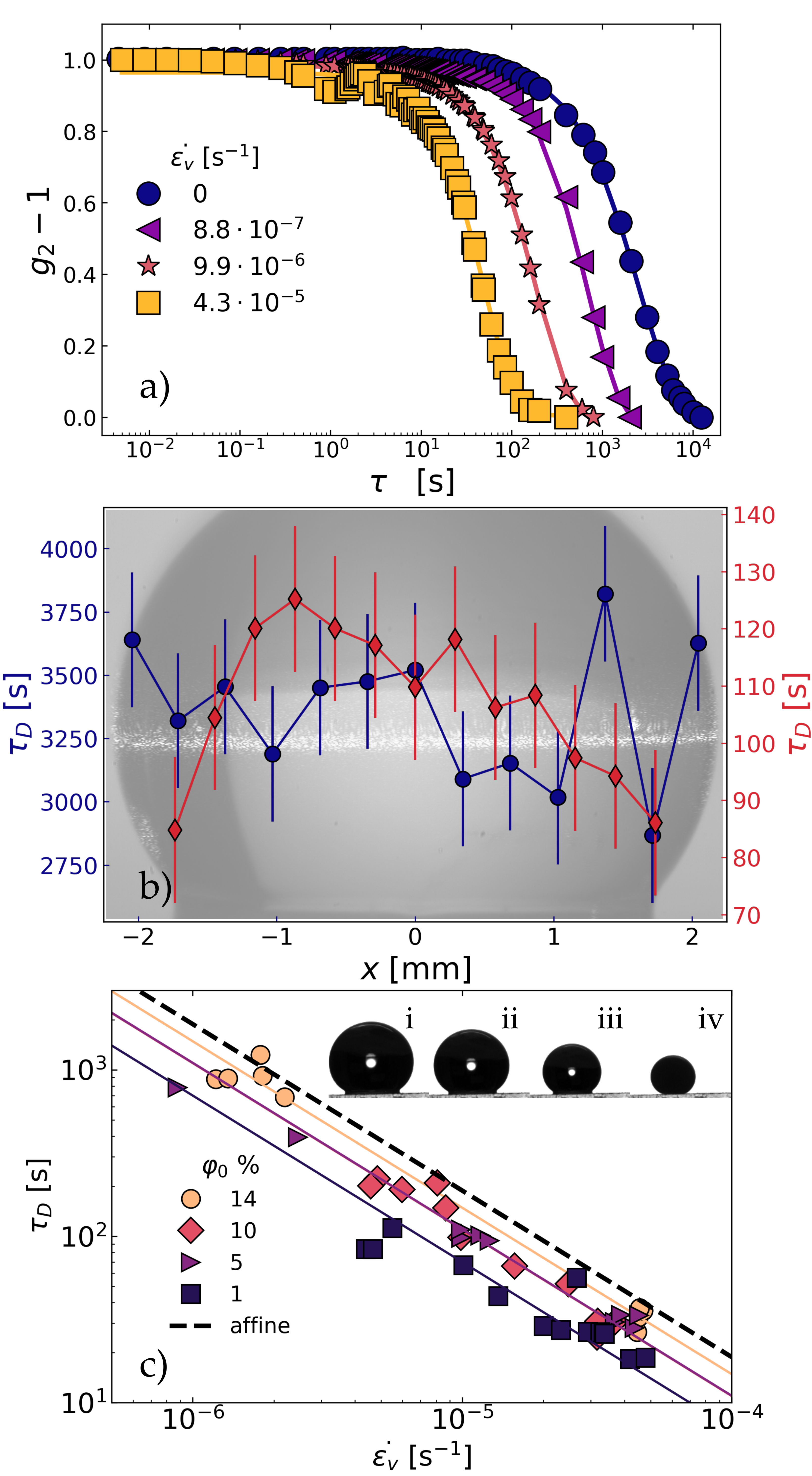}
\caption{\label{fig:DLS}a) Correlation functions measured in the center of a bead of gel prepared at $\varphi_0=5\%$, in a quiescent state and when compressed at different rates $\dot{\varepsilon_v}$. Symbols are data and solid lines are fits to a compressed exponential decay. b) \RB{Decay time $\tau_D$ }, as a function of $x$, for a quiescent (blue circles) and a drying (red diamonds) bead. Background: Image of a gel bead illuminated simultaneously by white and laser lights. c) $\dot{\varepsilon_v}$-dependence of the relaxation time $\tau_D$ measured at $x=0$, for samples prepared at different initial volume fractions as indicated. The dashed line is the theoretical expectation for a homogeneous elastic sphere undergoing a purely affine compression~\cite{milani2024space}. Thin solid lines are  fits to the data for each $\varphi_0$ with a power law with a fixed exponent $-1$. Inset: snapshots of a drying bead taken at times $0,50,150,200$ min for panels $i$ to $iv$ respectively. The scale bar represents $1$ mm.}
\end{figure}

To gain insight into gel compression at the microscopic scale, we use a custom-made dynamic light scattering setup based on Photon Correlation Imaging~\cite{duri2009resolving}, specifically designed to probe microscopic dynamics with time- and space-resolution in spherical samples~\cite{milani2024space,milani2023double}. In brief, a laser beam illuminates the bead along a diameter (we periodically adjust the vertical bead position during drying to ensure that the laser always hits the bead along a diameter~\cite{milani2024space}) and a CMOS camera takes images of the speckle pattern generated by the scattered light, see the background of Fig.~\ref{fig:DLS}b. We measure the decrease of the bead radius $R$ due to drying and the evolution of the speckle pattern. The microscopic dynamics are quantified by calculating, for different positions $x$ along the bead diameter, $g_2(\tau,q)-1$, the intensity autocorrelation of the temporal fluctuations of the light scattered at a wave vector $q = 22.2~\mu\mathrm{m}^{-1}$~\cite{milani2023double,milani2024space}. 

We plot in Fig.~\ref{fig:DLS}a correlation functions measured in the center of the bead, $x=0$, for a gel prepared at $\varphi_0=5\%$. In all cases $g_2-1$ decays to zero, indicative of the full relaxation of density fluctuations on length scales $\sim q^{-1} \approx 50$~nm. The quiescent sample (before drying) exhibits slow dynamics (time scale $\sim 5000$~s), which we attribute to the relaxation of internal stresses built up during gelation, in agreement with previous measurements for similar gels in the bulk~\cite{cipelletti2000universal, aime_microscopic_2018,
cipelletti2003universal,joshi2014dynamics}. During drying, the microscopic dynamics speed up and are faster for higher compression rates $\dot{\varepsilon_v}$.  Data are fitted with a compressed exponential function with mean relaxation time $\tau_D$ (see~\cite{SM} for details). We find that $\tau_D$ is independent of the position in the bead, for both quiescent and drying gels (see representative data in Fig.~\ref{fig:DLS}b), a strong indication of the absence of any significant dynamical inhomogeneity in the gel even when it is compressed. This finding is in agreement with recent observations on the syneresis of quasi-2D gels composed of stiff particles~\cite{wu2023spatially}.

Measurements performed at different RH and $\varphi_0$ can be rationalized by plotting $\tau_D$ \textit{vs} $\dot{\varepsilon_v}$, Fig.~\ref{fig:DLS}c. \footnote{note that plotting $\tau_D$ $\textit{vs}$ the instantaneous $\varepsilon_v$ or volume fraction increase does not yield master curves~\cite{SM}.} Overall, $\tau_D$ decreases as $\dot{\varepsilon_v}$ increases, showing that the macroscopic compression rate has a direct effect on the microscopic motion within the sample, similarly to the observations for other driven glassy and jammed materials~\cite{brambilla2011highly, milani2023double,lee2009direct,besseling2007three,varnik2006structural}. Indeed, the experimental data are very close \LC{to}, although systematically smaller \LC{than}, the theoretical expectation $\tau_D \propto \dot{\epsilon_v}^{-1}$ for a perfectly elastic sphere undergoing affine compression, where the prefactor has been computed as in~\cite{milani2024space} (dashed line in Fig.~\ref{fig:DLS}c). This, together with the global fastening of the dynamics, indicates that the stress applied on the surface of the sample by the receding drop propagates to the whole gel volume due to its elastic nature. Furthermore, the fact that experimental data lay below the dashed line indicates that the sample undergoes some plastic rearrangements during compression, which accelerate the microscopic dynamics up to a factor of three
as compared to affine predictions. This finding, consistent with recent 2D simulations but not tested so far experimentally, shows that isotropic compression of a colloidal gel induces non-affine deformations that are roughly constant during the whole compression process, as opposed to axial compression in which non-affine deformations increase in time~\cite{islam2021consolidation}. The existence of plastic rearrangements is also consistent with the macroscopic observation that drying is irreversible: compressed beads of colloidal gels do not re-swell in their solvent~\cite{SM}, unlike polymer gels~\cite{milani2024space}. By fitting the experimental data with a power law, $\tau_D = A \dot{\epsilon_v}^{-1}$, we find that the prefactor $A$ increases twofold  when $\varphi_0$ increases from $1$ to $14$ \%. Thus, compression-induced plasticity is enhanced in weak, tenuous gels.

\begin{figure}[h]
\includegraphics[width=1\columnwidth]{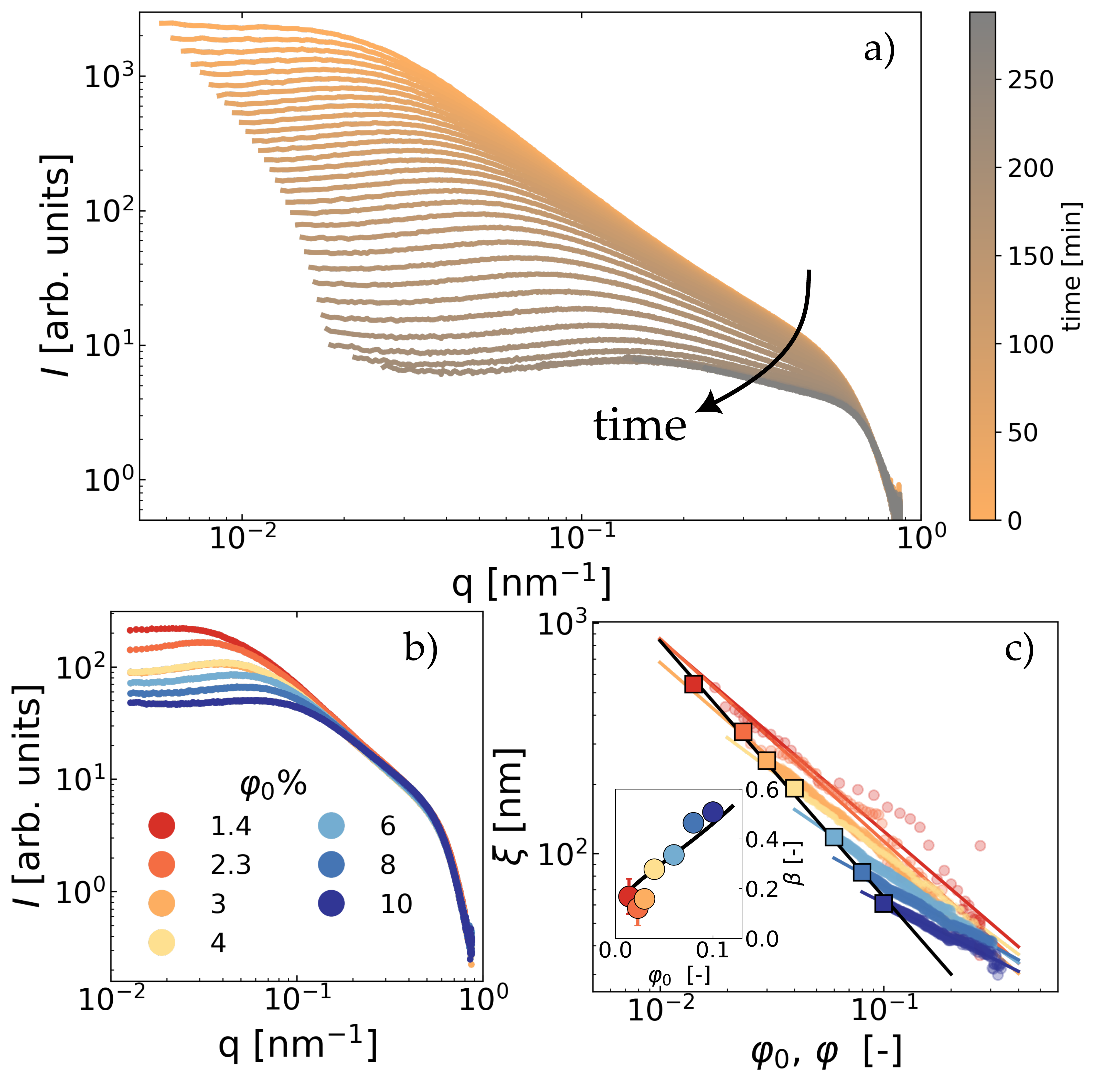}
\caption{\label{fig:Xray} a) Time evolution of the normalized scattered intensity, $I(q)$, during the drying of a bead of colloidal gel with $\varphi_0=3$ \%. b) $I(q)$  for gel beads dried to the same final volume fraction $\varphi_f=10$ \% but prepared at different initial volume fractions $\varphi_0$, as indicated by the labels. c) Small symbols: cutoff length for density fluctuations \textit{vs} $\varphi$ during drying. Each color corresponds to a different $\varphi_0$, same colors as in panel b). Large squares: $\xi$ \textit{vs} $\varphi_0$ for pristine gels before drying,  $t=0$. \LC{Solid lines:} power law fits to the data. \LC{Inset: $\beta$ coefficient \textit{vs} $\varphi_0$ (symbols) and fit (line), see text and \cite{SM} for details}.}
\end{figure}

To probe the structural evolution of the gel network resulting from plasticity , we use time-resolved synchrotron small-angle X-ray scattering (SAXS). The X-ray beam (size $200$ $\times$ $200$ $\mu$m$^2$) initially \LC{propagates along a bead diameter}, but as the bead shrinks during drying, it \LC{intersects the bead increasingly} closer to the pole. The measured scattered intensity \LC{$I$ is averaged over} the whole scattering volume; since the microscopic dynamics is constant along a bead diameter (Fig.~\ref{fig:DLS}b), we expect the structure to be spatially homogeneous as well during the whole drying process, as checked by scanning the microscopic structure at different positions~\cite{SM}. Moreover, from the comparison of data acquired with two different RH, we conclude that structural evolution does not depend on the compression rate~\cite{SM}. Hence, most data are acquired for beads that evaporate in open air, to minimize spurious scattering that would be induced by additional windows to enclose the sample.  

The scattering at time $0$, $I_0$, is related to the structure of a pristine colloidal gel~\cite{carpineti_spinodal-type_1992,j._bibette_kinetically_1992,carpineti1993transition,carpineti1995mass} and has the following specific features. The scattering of individual NPs is probed at a high scattering vector $q$ ($q>0.5$ nm$^{-1}$). At intermediate \LC{$q$, $I_0 \sim q^{-\alpha}$, with $\alpha \approx 2.0$ an estimate of} the fractal dimension of the NP clusters that form the network. Finally, $I_0$ plateaus at small $q$, indicating that, on a sufficiently large length scale, the gel structure becomes homogeneous. The cross-over between the two regimes occurs at a characteristic $q^{*}$ from which we extract the cutoff length for density fluctuations, $\xi = 2\pi/q^{*}$~\cite{SM}. As $\varphi_0$ increases, $\xi$ decreases from hundreds to a few tens of nanometers following a power law, $\xi \sim \varphi_0^{p}$ with \RBB{$p=-1.11 \pm 0.03$. One expects $p=-\frac{1}{3-d_f}$~
\cite{carpineti_spinodal-type_1992,j._bibette_kinetically_1992,carpineti1995mass}, yielding a fractal dimension $d_f=2.10 \pm 0.05$, in good agreement with the $\alpha$ exponent characterizing the power-law decrease of $I(q)$.}

We now describe the evolution of the bead microstructure during drying, with $\phi(t)$ \LC{obtained from the volume of the bead, which is imaged} concomitantly to SAXS \LC{measurements}~\cite{SM}. Beads prepared at various $\phi_0 \in [1.4\%,10\%]$ exhibit qualitatively similar behavior. As an illustration, Fig.~\ref{fig:Xray}a displays $I(q)$ for $\varphi_0 =3\%$ \footnote{The curves are plotted with different $q$ range according to the drying process. This is due to the spurious contributes coming from surface scattering }. Scattering curves are normalized so that they overlap in the high $q$ region
~\cite{SM}, thus accounting for the change of the scattering volume.  The measurement lasts about $300$ min and the final volume fraction typically reaches $30\%$. \LC{Remarkably, the same $d_f$ is kept throughout drying, while the cutoff length $\xi$}  increasingly departs from \LC{its initial value}, decreasing as the colloidal network is compressed and $\varphi$ \LC{grows.} For pristine gels, increasing $\varphi_0$ also induces a decrease in $\xi$ (large squares in Fig.~\ref{fig:Xray}c), suggesting that increasing $\varphi$ by drying or directly during sample preparation may lead to the same microscopic structure. To test this hypothesis, we compare the structure of several beads prepared at different $\varphi_0 \leq 10\%$ and subsequently dried to a final volume fraction $\varphi_f = 10\%$~\footnote{The gel prepared at $\varphi_0 = 10\%$ is kept pristine}. Remarkably, we find that, \RBB{despite having the same structure at high $q$,} the low $q$ intensity depends very strongly on the sample history, \RBB{indicating that the beads preserve a \textit{memory} of their pristine structure}.

\RBB{To better understand the structural evolution of the gels, we plot in Fig.~\ref{fig:Xray}c the cutoff length \textit{vs} the instantaneous volume fraction, for gels prepared at different $\varphi_0$.} \LC{During drying $\xi(\varphi)$ follows a power law, with an increasingly weaker dependence on $\varphi$ as the initial volume fraction increases. This, together with the fact that gels under compression retain the same $d_f$ as the pristine ones, suggests that $\xi(\varphi)$ may be written by generalizing the  expression for pristine gels:
\begin{equation}
\xi \sim \varphi^{-\frac{1-\beta}{3-d_f}}\,,
\label{eq:xi overlap}
\end{equation}
where the correction term $\beta$ vanishes for pristine gels. We find that $\beta$ increases with $\varphi_0$ from $0.1$ to $0.5$~\cite{SM}, which may be rationalized by noting that all $\xi({\phi})$ data seemingly extrapolate to the size of closely packed small clusters of about four particles~\cite{SM}.}

\begin{figure}[h]
\includegraphics[width=1\columnwidth]{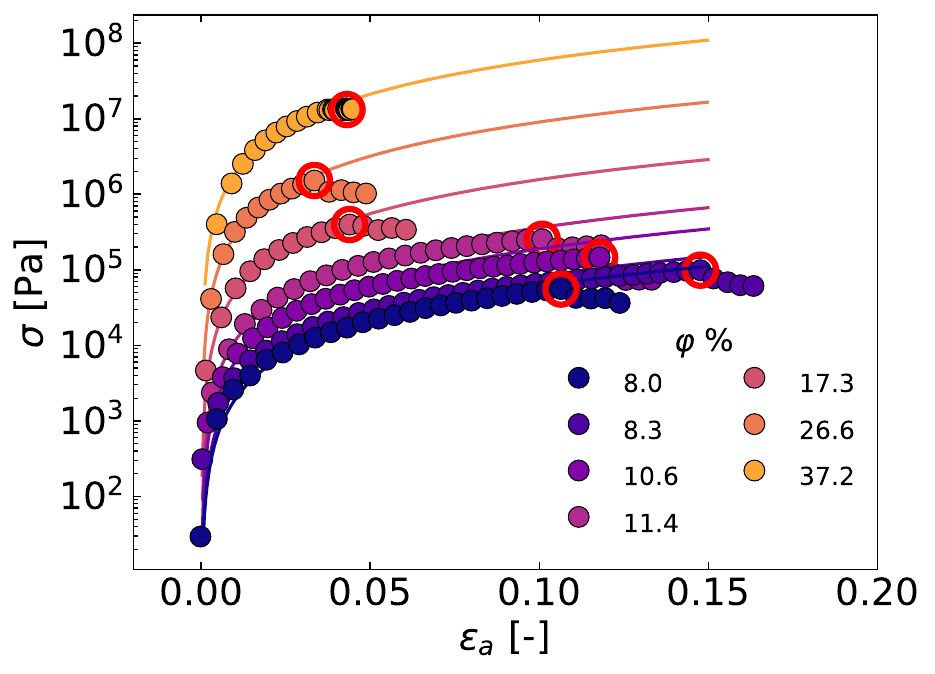}
\caption{\label{fig:force_distance} Stress \textit{vs} strain curves for gel beads prepared at the same $\varphi_0=8\%$, dried to the $\phi$ indicated by the label, and then submitted to uniaxial compression at a rate $\dot{\varepsilon_{\rm{a}}}= 0.1$~s$^{-1}$. The open red circles indicate the yield point. Symbols are data and lines are best fits using the Hertz contact model~\cite{SM}.}
\end{figure}

\begin{figure}[h]
\includegraphics[width=1\columnwidth]{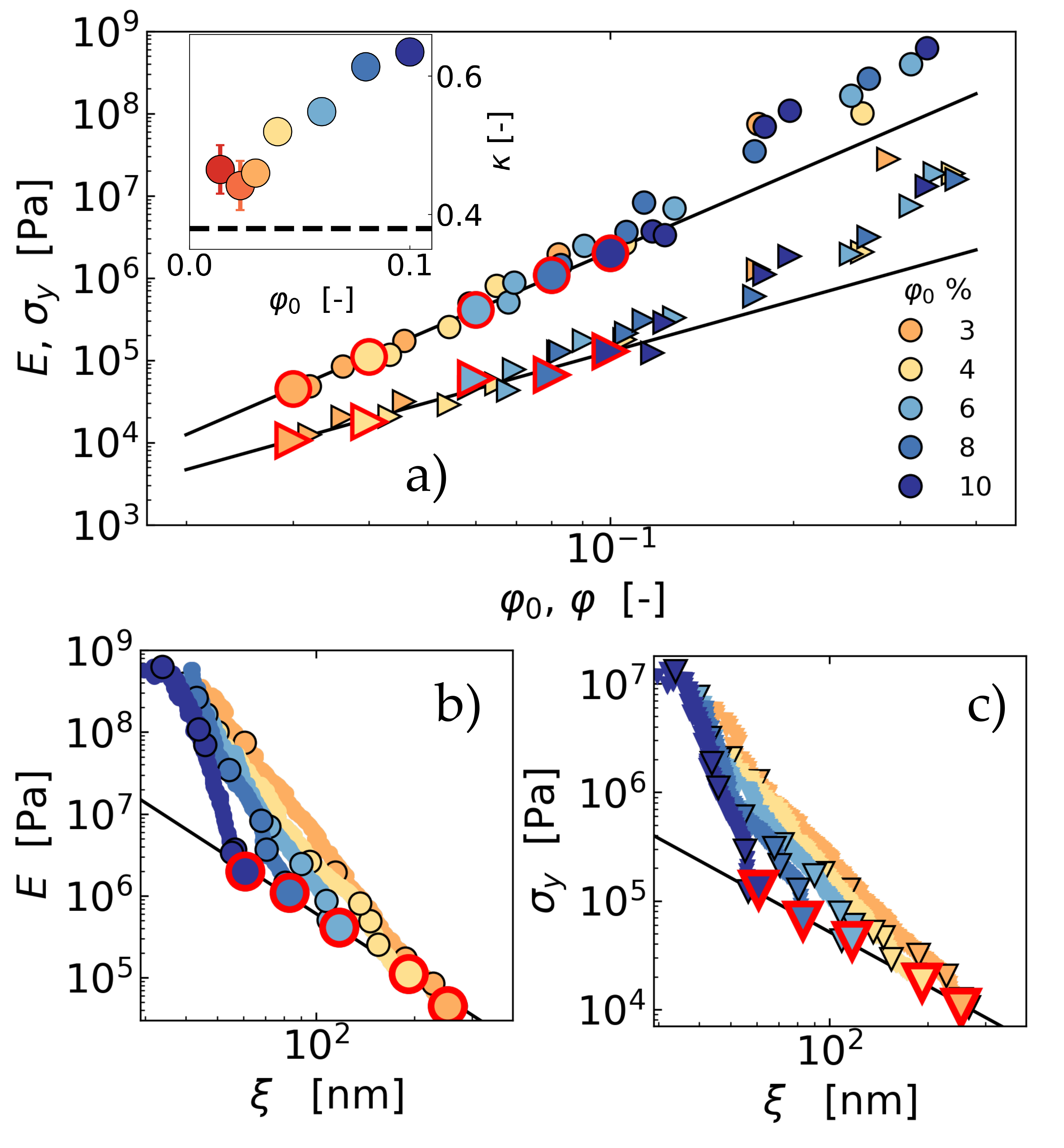} 
\caption{\label{fig:rheo}a) Young modulus, $E$ (circles) and yield stress,  $\sigma_y$ (triangles) \textit{vs} volume fraction. Large symbols with red edge correspond to pristine gels measured before drying, small symbols are obtained at different times during the drying process, for beads prepared at different initial volume fractions $\varphi_0$ as indicated by the labels. \LC{Inset: $\kappa$ coefficient controlling the relative importance of stretching and bending modes \textit{vs} $\phi_0$, for pristine (dashed line) and compressed (symbols) gels}. b,c): $E$ and $\sigma_y$ \textit{vs} the cutoff size of density fluctuations. Lines are power law fits for pristine gels with exponents $m=2.59\pm0.06$ and  $n=1.65\pm0.06$ for b) and c), respectively.}
\end{figure}

One key aspect of pristine fractal colloidal gels is the link between their microscopic structure and mechanical properties~\cite{shih1990scaling,bantawa2023hidden,dages2022interpenetration,Ste}. \LC{In this framework, the Young modulus $E \sim K/\xi$, where the spring constant $K$ is the cluster stiffness. This corresponds to the \textit{strong-link} regime defined in the pioneered work of Shih~\textit{et al.}~\cite{shih1990scaling}, where $K$ is expected to vary with $\xi$ and hence $\varphi$ ~\cite{shih1990scaling,de1994elasticity,mellema2002categorization}.} We probe the mechanical properties of beads prepared at $3\% \leq \varphi_0 \leq 10\%$ and dried to reach different $\phi$ values\LC{, using} uniaxial compression tests to measure \RB{engineering} stress ($\sigma$)-strain ($\varepsilon_{\rm{a}}$) curves~\cite{SM}. 
Figure~\ref{fig:force_distance} displays representative \LC{data} for beads prepared at the same initial $\varphi_0=8\%$ and dried to various degrees to reach volume fractions $\varphi$ up to $37.4$ \%.  In all cases, $\sigma$ increases with $\varepsilon_{\rm{a}}$ reaching a maximum value identified as the yield stress $\sigma_y$ (red open circles), before dropping as the gel yields. 
We fit the low-strain data \RB{($\varepsilon_{\rm{a}}<5\%$)} to the Hertz contact model to determine $E$~\cite{SM}, and plot $E$ and $\sigma_y$ \LC{\textit{vs} volume fraction} in Fig.~\ref{fig:rheo}a. 

Data for pristine gels evolve as power laws:  $E \sim \varphi_0^{\gamma}$, $\sigma_y \sim \varphi_0^{\delta}$, with exponents $\gamma=3.18\pm 0.02$ and $\delta=2.05\pm 0.03$. \LC{ These values are typical of gels in the strong link regime~\cite{shih1990scaling,de1994elasticity,mellema2002categorization},} \LC{ where the cluster elastic constant reads
\begin{equation}
K \sim \varphi_0^{(2\kappa + d_b)/(3 - d_f)}\,,
\label{eq:K}
\end{equation}
with $d_b = 1.1$ the backbone dimension~\cite{shih1990scaling}. The parameter $\kappa$ describes the relative weight of contributions to elasticity  arising from bending ($\kappa=1$) or stretching ($\kappa=0$) deformations of the gel strands, respectively~\cite{mellema2002categorization}. Using $K \sim  \xi E$ and  combining Eqs.~\ref{eq:xi overlap} and \ref{eq:K}, for the pristine gels ($\beta=0$, $\phi \equiv \phi_0$) one finds $K \sim \xi E \sim \phi_0^{\nu_0}$, with $\nu_0 = \gamma + \frac{\-1}{3-d_f}$, leading to $\kappa \equiv \kappa_0= \left[ (3-d_f)\gamma-1-d_b\right]/2 = 0.38$, indicating that both bending and stretching modes contribute to the cluster elasticity, but the latter dominate. Finally, the above scaling laws imply a unique bijective relationship between mechanical and structural quantities: $E\sim \xi^{-m}$, $m=2.59\pm0.06$, and $\sigma_y \sim \xi^{-n}$, $n=1.65\pm0.06$, see Fig.~\ref{fig:rheo}b,c.} 

During drying, $E$ and $\sigma_y$ increase monotonically with $\varphi$, indicating a stiffening of the gels (Fig.~\ref{fig:rheo}a), similar to the gel consolidation stage reached during uniaxial compression~\cite{roy2016universality}. 
Remarkably, data for all compressed gels perfectly overlap with the power law established for pristine gels. Only for $\varphi>10\%$ do measurements for compressed gels depart from the extrapolated trend of pristine systems, presumably because of the breakdown of the description in terms of assembly of fractal clusters. \LC{While the mechanical properties of compressed gels follow the same $\phi$ scaling as those of pristine ones, we remind that their structural length $\xi$ depends on $\phi$ via distinct power laws, according to the initial composition (Fig.~\ref{fig:Xray}c and Eq.~\ref{eq:xi overlap}). As a consequence, the bijective relation seen for pristine gels (lines in Fig. 4b,c) breaks down for compressed gels. Remarkably, this makes accessible a new 3D region in the $(E,\sigma_y,\xi)$ parameter space, whose 2D projections are shown in Fig.~\ref{fig:rheo}b,c.} In line with microscopic dynamic observations that show a behavior closer to that of a perfectly elastic body as $\varphi_0$ increases, we find here that the $E$ \textit{vs} $\xi$ curve is steeper when $\varphi_0$ is higher.

 \LC{To rationalize the key features of the elasticity of compressed gels, we propose that the spring constant $K$ of the elastic units still obeys Eq.~\ref{eq:K}, but with a modified balance between bending and stretching contributions, depending on drying. Following the same steps as for pristine gels, for the general case $\beta > 0$ one finds $\kappa = \left[ (3-d_f)\gamma + \beta -1-d_b\right]/2=\kappa_0+\beta/2$, with $\kappa_0=0.38$ for pristine gels~\cite{SM}. In our experiments, $\beta$ grows from $0.1$ to $0.5$ with increasing $\phi_0$, corresponding to $\kappa$ values from $0.43$ to $0.63$, see inset of Fig.~\ref{fig:rheo}a. The growth of $\kappa$ suggests that plasticity stiffens the network by increasing the bending-to-stretching ratio: as the gel compactifies due to drying, new bonds form, enhancing the bending rigidity of the gel strands.} 

The non-bijective relationship between linear elasticity and structural length scale uncovered in this work shows how compressed gels, despite exhibiting structure and mechanical properties similar to pristine ones, undergo a profound change in the origin of their elasticity. \LC{In future work, the relation between structure and elasticity could be further explored  by varying the microscopic structure of the gels, e.g. by altering the aggregation dynamics, the range and depth of the interparticle potential, or the compressibility of the primary particles. Finally, the results presented here suggest that, quite generally}, isotropic compression may be a promising way to decouple mechanical properties from microstructure, with implications in the design of amorphous solids  with unique, optimized microstructure and mechanical properties.

\section*{Data Availability}
The data that support the findings of this study are available on Zenodo~\cite{zenodo}.

\begin{acknowledgments}
We thank J. Barbat,  G. Prévot, and R. Jelinek for help with instrumentation, and T. Bizien and W. Chèvremont for help with the Synchrotron measurements at Soleil and ESRF, respectively. Discussions with S. Aime, E. Del Gado, T. Divoux, \LC{M. Lenz, D. Richard} are gratefully acknowledged.  We acknowledge financial support from the French Agence Nationale de la Recherche (ANR) (Grant No. ANR-19-CE06-0030-02, BOGUS), from the GDR2019 CNRS/INRAE “Solliciter LA Matière Molle” (SLAMM), and from Synchrotron Soleil (beamline SWING) and ESRF (beamline ID2). LC acknowledges support from the Institut Universitaire de France. AM is a postdoctoral researcher supported by the French CNES (Centre National d’Etudes Spatiales). 

\end{acknowledgments}




\nocite{zenodo}

\bibliography{apssamp}

\end{document}